\documentclass[12pt,preprint]{aastex}
\usepackage{emulateapj5}
\usepackage{apjfonts}
\usepackage{epsfig}
\usepackage{natbib}

\newcommand{\aox}{\ensuremath{\alpha_{\mathrm{ox}}}}
\newcommand{\cs}{\emph{c}}

\newcommand{\chandra}{\emph{Chandra}}

\newcommand{\etal}{et al.}

\newcommand{\flamb}{ergs s$^{-1}$ cm$^{-2}$ \AA$^{-1}$}
\newcommand{\flux}{ergs s$^{-1}$ cm$^{-2}$}

\newcommand{\feii}{\ion{Fe}{2}}

\def\gtrsim{\mathrel{\hbox{\rlap{\hbox{\lower4pt\hbox{$\sim$}}}\hbox{\raise2pt\hbox{$>$}}}}}

\newcommand{\fwha}{\ensuremath{\mathrm{FWHM}_\mathrm{H{\alpha}}}}

\newcommand{\halpha}{H\ensuremath{\alpha}}

\newcommand{\hbeta}{H\ensuremath{\beta}}

\newcommand{\hst}{\emph{HST}}

\newcommand{\kms}{km~s\ensuremath{^{-1}}}

\newcommand{\lf}{\ensuremath{L_{\rm{5100 \AA}}}}
\newcommand{\lha}{\ensuremath{L_{\mathrm{H{\alpha}}}}}

\newcommand{\lledd}{\ensuremath{L_{\mathrm{bol}}/L{\mathrm{_{Edd}}}}}

\newcommand{\lsun}{\ensuremath{L_{\odot}}}
\newcommand{\lum}{ergs s$^{-1}$}

\newcommand{\loiii}{\ensuremath{L_{\mathrm{[O {\tiny III}]}}}}

\newcommand{\mbh}{\ensuremath{M_\mathrm{BH}}}

\newcommand{\msigma}{\ensuremath{M_{\mathrm{BH}}-\sigmastar}}
\newcommand{\msun}{\ensuremath{M_{\odot}}}

\newcommand{\nii}{[\ion{N}{2}]}
\newcommand{\oi}{[\ion{O}{1}]}
\newcommand{\oii}{[\ion{O}{2}]}
\newcommand{\oiii}{[\ion{O}{3}]}

\newcommand{\psix}{\ensuremath{P_{\mathrm{6cm}}}}
\newcommand{\rosat}{\emph{ROSAT}}

\newcommand{\rkel}{{\emph R}}

\newcommand{\sii}{[\ion{S}{2}]}

\newcommand{\sigmastar}{\ensuremath{\sigma_{\ast}}}

\newcommand{\spitzer}{\emph{Spitzer}}

\newcommand{\xmm}{{\it XMM-Newton}}
\newcommand{\whz}{W~Hz$^{-1}$}

\newcommand{\zw}{1~Zw~{\small I}}

\def\lax{{$\mathrel{\hbox{\rlap{\hbox{\lower4pt\hbox{$\sim$}}}\hbox{$<$}}}$}}
\def\gax{{$\mathrel{\hbox{\rlap{\hbox{\lower4pt\hbox{$\sim$}}}\hbox{$>$}}}$}}

\slugcomment{To appear in {\it The Astrophysical Journal}.}
\shorttitle{Low-mass Black Holes}
\shortauthors{GREENE \& HO}

\begin{document}

\title{A New Sample of Low-mass Black Holes in Active Galaxies}

\author{Jenny E. Greene\altaffilmark{1}}
\affil{Department of Astrophysical Sciences, Princeton University, 
Princeton, NJ}
\altaffiltext{1}{Hubble Fellow and Princeton-Carnegie Fellow.}

\author{Luis C. Ho}
\affil{The Observatories of the Carnegie Institution of Washington,
813 Santa Barbara St., Pasadena, CA 91101}

\begin{abstract}

We present an expanded sample of low-mass black holes (BHs) found in
galactic nuclei.  Using standard virial mass techniques to estimate BH
masses, we select from the Fourth Data Release of the Sloan Digital
Sky Survey all broad-line active galaxies with masses $<2 \times
10^6$~\msun.  BHs in this mass regime provide unique tests of the
relationship between BHs and galaxies, since their late-type galaxy
hosts do not necessarily contain classical bulges.  Furthermore, they
provide observational analogs of primordial seed BHs and are expected,
when merging, to provide strong gravitational signals for future
detectors such as \emph{LISA}.  From our preliminary sample of 19, we
have increased the total sample by an order of magnitude to 174, as
well as an additional 55 (less secure) candidates.  The sample has a
median BH mass of $\langle$\mbh$\rangle = 1.3 \times 10^6$~\msun, and
in general the objects are radiating at high fractions of their
Eddington limits. We investigate the broad spectral properties of the
sample; 55 are detected by \rosat, with soft X-ray luminosities in the
range $10^{40}$ to $7 \times 10^{43}$ \lum.  Much like the preliminary
sample, these objects are predominantly radio-quiet (\rkel\ $\equiv
f_{\rm 6 cm}/f_{\rm 4400 \AA} < 10$), but 11 objects are detected at
20 cm, with radio powers ($10^{21}-10^{23}$ \whz) that may arise from
either star formation or nuclear activity; only $1 \%$ of the sample
is radio-loud.  We further confirm that, with $\langle M_g \rangle =
-19.3$ and $\langle g-r \rangle = 0.7$ mag, the host galaxies are
low-mass, late-type systems.  At least $40\%$ show disk-like
morphologies, and the combination of host galaxy colors and
higher-order Balmer absorption lines indicate intermediate-age stellar
populations in a subset of the sample.

\end{abstract}

\keywords{galaxies: active --- galaxies: nuclei --- galaxies: Seyfert} 

\section{Introduction}

There are many strong observational connections between galaxy bulges
and central supermassive black holes (BHs). BH masses correlate
surprisingly tightly with bulge properties, including luminosity
(e.g.,~Marconi \& Hunt 2003) and stellar velocity dispersion (\msigma;
Ferrarese \& Merritt 2000; Gebhardt \etal\ 2000a; Tremaine \etal\
2002).  Also active galactic nuclei (AGNs) in the local Universe are
predominantly found in massive, bulge-dominated galaxies (Ho \etal\
1997b; Kauffmann \etal\ 2003).  Perhaps the formation of BHs and
bulges are related, in which case we might not expect bulgeless
galaxies to obey scaling relations between BHs and galaxies, or even
necessarily host a central BH.  Indeed, dynamical study of the
bulgeless spiral galaxy M33 places strong limits on the presence of a
dark central massive object (Gebhardt \etal\ 2001), while any BH in
the nucleus of the dwarf spheroidal galaxy NGC 205 has been shown to
lie below the low-mass extrapolation of the \msigma\ relation (Valluri
\etal\ 2005).  On the other hand, the M31 globular cluster G1 shows
dynamical and radiative evidence for a central BH
(Gebhardt \etal\ 2002, 2005; Pooley \& Rappaport 2006; Ulvestad \etal\
2007).  However, the mixed stellar population and high degree of
rotational support in this massive cluster suggest that G1 is
actually the nucleus of a tidally stripped dwarf galaxy (e.g.,~Meylan
\etal\ 2001).  Thus, dynamical studies present an ambiguous verdict on
the presence of nuclear BHs in dwarf stellar systems.
Unfortunately, it is not currently feasible to spatially resolve the
gravitational sphere of influence of a $\sim 10^5$~\msun\ BH outside
the Local Group, in order to search for BHs in more dwarf stellar
systems using dynamical methods.

Although they are difficult to find, the occupation fraction of
nuclear BHs in dwarf systems and the space density of low-mass BHs are
of considerable interest.  Apart from furnishing additional insight
into the possible origin of the \msigma\ relation, low-mass BHs
provide low-redshift counterparts to the primordial seed BHs; the
low-mass cut-off in the BH mass function today provides a constraint
on the mass function of seed BHs.  The merging of BHs in this mass
range is expected to provide a strong signal for the gravitational
wave experiment \emph{LISA} (e.g.,~Hughes 2002).  Furthermore,
gravitational radiation recoil is expected to impart velocities to BH
merger remnants that exceed the escape velocities of dwarf galaxies
(e.g.,~Favata \etal\ 2004; Merritt \etal\ 2004).  This effect alone
might decrease the occupation fraction of BHs in dwarf galaxies.

In the absence of detectable dynamical signatures, we are forced to
rely on less direct evidence for the presence of nuclear BHs, namely
AGN activity.  In fact, there are two well-studied AGNs in dwarf
galaxies: NGC 4395 is a bulgeless spiral galaxy (Filippenko \& Sargent
1989), while POX 52 (Kunth \etal\ 1987) is a dwarf spheroidal galaxy
(Barth \etal\ 2004)\footnote{Such objects are commonly referred to as
dwarf elliptical galaxies in the literature.  However, because their
structure is quite different from that of classical elliptical
galaxies, we prefer to refer to them as dwarf spheroidal systems.}.
BH mass determinations are far less certain in the absence of
dynamical constraints, but a variety of techniques yield BH masses for
each object of $\sim 10^5$~\msun\ (Filippenko \& Ho 2003; Barth \etal\
2004; Peterson \etal\ 2005).  Greene \& Ho (2004; GH hereafter)
performed a systematic search for AGNs with 

\hskip -0.15in
\psfig{file=tablebasic_sh.epsi,width=0.5\textwidth,keepaspectratio=true,angle=0}
\vskip 4mm
\noindent
BH masses $<10^6$~\msun, using the First Data Release of the Sloan
Digital Sky Survey (SDSS; York \etal\ 2000) and recovered 19 objects
in that preliminary search.  Remarkably, follow-up spectroscopy using
ESI on Keck suggests that these sources, as well as NGC 4395, POX 52,
and G1, are consistent with the low-mass extrapolation of the \msigma\
relation (Barth \etal\ 2005).  At the same time, many of the host
galaxies are late-type galaxies without classical bulges
(J.~E.~Greene, et al. in preparation).  Clearly, more objects are
needed to both confirm this preliminary result, and also to constrain
the structural parameters of the host galaxies.  To this end, we have
repeated our analysis using the Fourth Data Release of the SDSS (DR4;
Adelman-McCarthy \etal\ 2006), and present the updated sample here.

Our goal is to study the properties of the lowest-mass BHs that are
identifiable with the SDSS.  The upper mass limit on this sample is
somewhat arbitrary.  Within the inactive sample of galaxies with
dynamical BH masses, the BH in M32, \mbh=$(2.5^{+3.0}_{-2.0}) \times
10^6$~\msun, has the lowest mass (Tremaine \etal\ 2002; excluding the
Milky Way, whose modern mass is $3.5 \times 10^6$~\msun; e.g.,~Ghez
\etal\ 2005; Eisenhauer \etal\ 2005).  We have adopted $2 \times
10^6$~\msun\ as the upper limit for our search.  Because dynamical
methods cannot extend into this regime, we previously had constraints
on neither the existence of lower-mass BHs nor on whether they obey
similar scaling relations with galaxy bulge properties as their
high-mass cousins.  Furthermore, while there are good reasons to
believe that not every dwarf stellar system contains a BH, we do not
know at what mass the occupation fraction departs from unity.

Throughout we assume the following cosmological parameters to calculate
distances: $H_0 = 100~h = 71$~\kms~Mpc$^{-1}$, $\Omega_{m} = 0.27$,
and $\Omega_{\Lambda} = 0.75$ (Spergel \etal\ 2003).

\section{Sample Selection and Analysis}

\begin{figure*}
\vbox{ 
\vskip -0.1truein
\hskip 0.in
\psfig{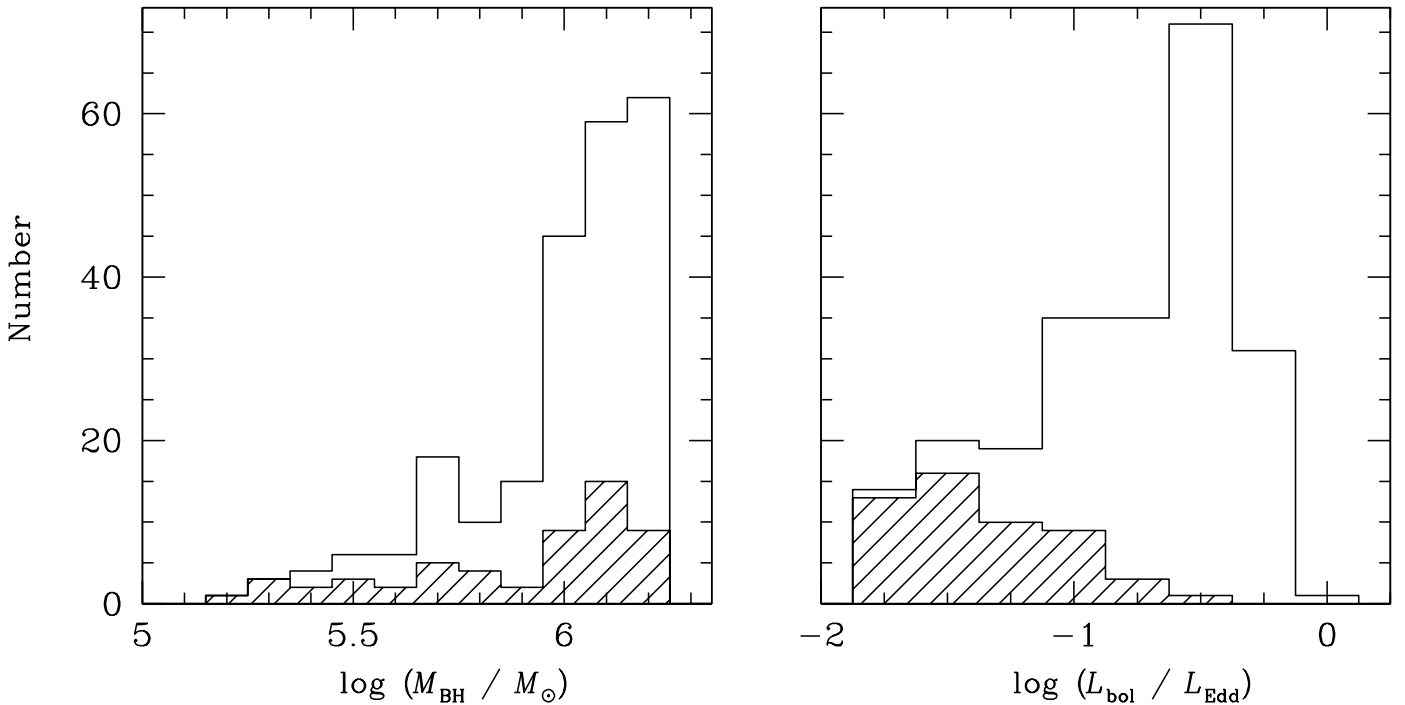}
}
\vskip -0mm
\figcaption[]{
Distributions of ``virial'' BH mass and Eddington ratio for the entire
sample.  \mbh\ is calculated from \lha\ and \fwha,
using the formalism of Greene \& Ho 2005b.  The Eddington ratio is
derived assuming an average bolometric correction of $L_{\rm{bol}} =
2.34 \times 10^{44} (L_{\rm H\alpha} / 10^{42})^{0.86}$ \lum\ (see
text).  Filled histograms represent objects below the 
detection threshold defined in Greene \& Ho 2007b (the {\it c} sample).
\label{masshist}}
\end{figure*}

Our low-mass BHs are selected from the sample of broad-line AGNs with
$z<0.352$ described in detail by Greene \& Ho (2007b) and briefly
reviewed here for completeness.  We begin with DR4 of the SDSS and
search for all AGNs with broad \halpha, where ``broad'' in this
context means a significant extra component relative to the
narrow-line profile (based on the \sii~$\lambda \lambda 6716,~6731$
doublet; e.g.,~Ho \etal\ 1997c).  Continuum subtraction, which is
crucial to uncover low-contrast broad lines, is performed with the
principle component analysis method developed by Hao \etal\ (2005).
Our subsequent selection technique comprises a two-step procedure: we
first select objects with high root-mean-square (rms) deviations above
the continuum in the region potentially containing broad \halpha, and
then perform more detailed profile fitting to isolate those objects
with broad \halpha\ profiles.  Unfortunately, this selection process
results in many sources of such low \halpha\ luminosity that (1) their
nature is ambiguous and (2) our ability to recover a reliable BH mass
is severely compromised.  Based on simulations, we thus impose a
combined rms-weighted flux and equivalent width (EW) cut designed to
minimize spurious detections.  The resulting sample comprises 8435
objects.

Since we cannot use stellar dynamical methods to measure BH masses, we
instead use the photoionized broad-line region (BLR) gas as a
dynamical tracer of the BH mass.  While the BLR velocity dispersion is
derived from the line width, the BLR radius is inferred from the AGN
luminosity (in this case \halpha; Greene \& Ho 2005b).  The so-called
radius-luminosity relation is derived from reverberation mapping of
$\sim 30$ AGNs, for which radii are measured based on the delay
between variations in the AGN photoionizing continuum and BLR line
emission (e.g.,~Kaspi \etal\ 2005; Greene \& Ho 2005b; Bentz \etal\
2006).  The ``virial'' BH mass is simply \mbh=$fR\upsilon^2/G$, where
$f$ is a scaling factor that accounts for the unknown geometry of the
BLR, assumed here to be spherical ($f$=0.75; Netzer 1990).  Although
BH masses derived from reverberation mapping are susceptible to large
systematic errors (due to uncertainties in the BLR geometry and
kinematics; e.g.,~Krolik 2001), remarkably they have been shown to
agree with \sigmastar, in those cases for which \sigmastar\ can be
measured.  The measured scatter is $\sim 0.3$ dex (Gebhardt \etal\
2000b; Ferrarese \etal\ 2001; Onken \etal\ 2004; Nelson \etal\ 2004).  
Virial masses are less direct, since they rely on the
radius-luminosity relation; in the largest such comparison to date,
Greene \& Ho (2006) find a scatter of $0.4$ dex of single-epoch virial
masses about the \msigma\ relation.  Our BH mass estimator is
discussed in detail in the Appendix.  Here we simply note that Greene
\& Ho (2007b) used an old version of the radius-luminosity relation
slope, while here we have updated the value to that found by Bentz
\etal\ (2006).  In general, the derived masses increase 
by $\sim 0.3$ dex.

Selecting those broad-line AGNs with masses $<2 \times 10^6$~\msun\
results in a sample of 174 objects, which are the main subject of this
paper (Fig. 1).  In addition, some (55) of the objects below the
detection threshold are nevertheless interesting candidates.  These
objects have been selected by manual inspection of the spectra, and
thus comprise a biased and incomplete sample.  We worry, as explained
above, that our BH masses are not stable for these \halpha\
luminosities, but we deem these objects worthy of follow-up
spectroscopy.  In what follows we will refer to these objects as the
\cs\ (``candidate'') sample, and we will present all trends including
and excluding these objects.  Basic properties of the sample are
summarized in Table 1.

\subsection{\feii\ Fitting}

Our modeling of the \hbeta\ region warrants some additional
discussion.  While the continua of our spectra, in general, are
dominated by stellar light, in some cases there is an additional
``pseudo-continuum'' component contributed by broad \feii\ multiplets.
The \feii\ emission extends over the entire optical and UV spectrum,
but is especially troublesome in the wavelength ranges 4400--4800 \AA\
and 5150--5500 \AA.  In principle, both the \hbeta\ and \oiii
~$\lambda \lambda$4959, 5007 fits can be severely compromised when the
\feii\ component is ignored (see Fig. 2).  We follow standard
procedure and model the \feii\ emission with an empirical template
(e.g.,~Boroson \& Green 1992; Greene \& Ho 2005b).  Since many of our
sources also require substantial galaxy subtraction, the \feii,
\hbeta, and \oiii\ lines are modeled simultaneously in the
galaxy-continuum--subtracted spectrum, as described below.

In addition to an \feii\ template, our model includes narrow \oiii,
and narrow and broad \hbeta.  In the standard way, all relative
wavelengths for central narrow components are fixed to laboratory
values, and the narrow \hbeta\ flux is fixed to be no more than 1/3.1
of the narrow component of \halpha\ (Case B$^{\prime}$ recombination;
Halpern \& Steiner 1983; Gaskell \& Ferland 1984).  The \oiii\ lines
are fit simultaneously with a core and a wing component, following
Greene \& Ho (2005a), with a relative strength of 2.96, while the
narrow \hbeta\ line is modeled with a single Gaussian due to its
typical low signal-to-noise ratio.  

Usually, \feii\ templates are derived from \zw, a strong \feii\ source
with relatively narrow broad lines.  With a broad-line width of
FWHM~$=1240$ \kms (Boroson \& Green 1992), \zw\ is significantly
broader than many of the objects in our sample.  For this reason, we
are compelled to build our own \feii\ template from the SDSS data.  
Using a high S/N spectrum of SDSS J155909.62$+$350147.4, whose \fwha\ 
is 860 \kms, we model \hbeta+\oiii\ using the 4750--5100 \AA\ region with
no \feii\ template included (Fig. 2{\it a}), and subtract this model.
We further mask the regions 4840--4870 \AA\ and 5010--5018 \AA\ (note
these wavelengths are in vacuum), since these regions are dominated by
noise in the subtraction.  While the resulting template is similar to
\zw\ (Fig. 2{\it b}), it is clear that we do not recover the shape of
the 5000 \AA\ feature perfectly, and that there are slight differences
in continuum slope between the two templates.  To explore the
importance of the former problem, we have repeated our fitting
procedure using templates made with various masking regions, and find
no difference in the derived fit parameters.  The continuum shape, on
the other hand, remains a systematic uncertainty inherent to our
method; there is some degeneracy between the shape of the continuum
and the amplitude of \feii\ contamination, leading to uncertainties in
the overall amplitude of \feii.  Nevertheless, our measured
\hbeta+\oiii\ fits are robust to this uncertainty.  When we fix the
\feii\ template to values ranging from one-half to twice the best-fit
value, we find $<1\%$ changes in all other measured parameters.
Presumably the \oiii\ is sufficiently resolved from the \feii\
component at 5000 \AA\ so that small errors in the latter do not
strongly impact our model of the former.

\subsection{Comparison with Greene \& Ho (2004)}

Let us examine how the original GH sample compares with this new
larger sample.  Objects from the original paper will be referenced
using their identification number in that paper (e.g., GH02).  There
are no significant differences between the two works, but there are
some issues worth noting.  In general, we have improved our selection
procedure (see Greene \& Ho 2007b), so that our sample from DR1 is
close to two and a half times larger than the original sample.  We are
now using the \halpha\ rather than the \lf\ luminosity to derive the
BLR radius, and we are using the updated radius-luminosity relation
slope of Bentz \etal\ (2006).  Only one (GH19) of the GH objects is
not included in the final DR4 broad-line AGN sample of Greene \& Ho
(2007b), although GH07, GH11, GH15, GH16, and GH18 now have masses
above our mass boundary of $2 \times 10^6$~\msun.  Finally, the
original GH sample excluded galaxies with significant galaxy
contamination, due to the fear that these objects would comprise more
massive BHs with low-contrast broad lines that were rendered
undetectable by the large galaxy luminosity.  In this work we have not
explicitly included such a cut, but our EW threshold is operationally
similar.  While some of the objects in the \cs\ sample may consist of
high-mass interlopers, 

\hskip -0.1in
\psfig{file=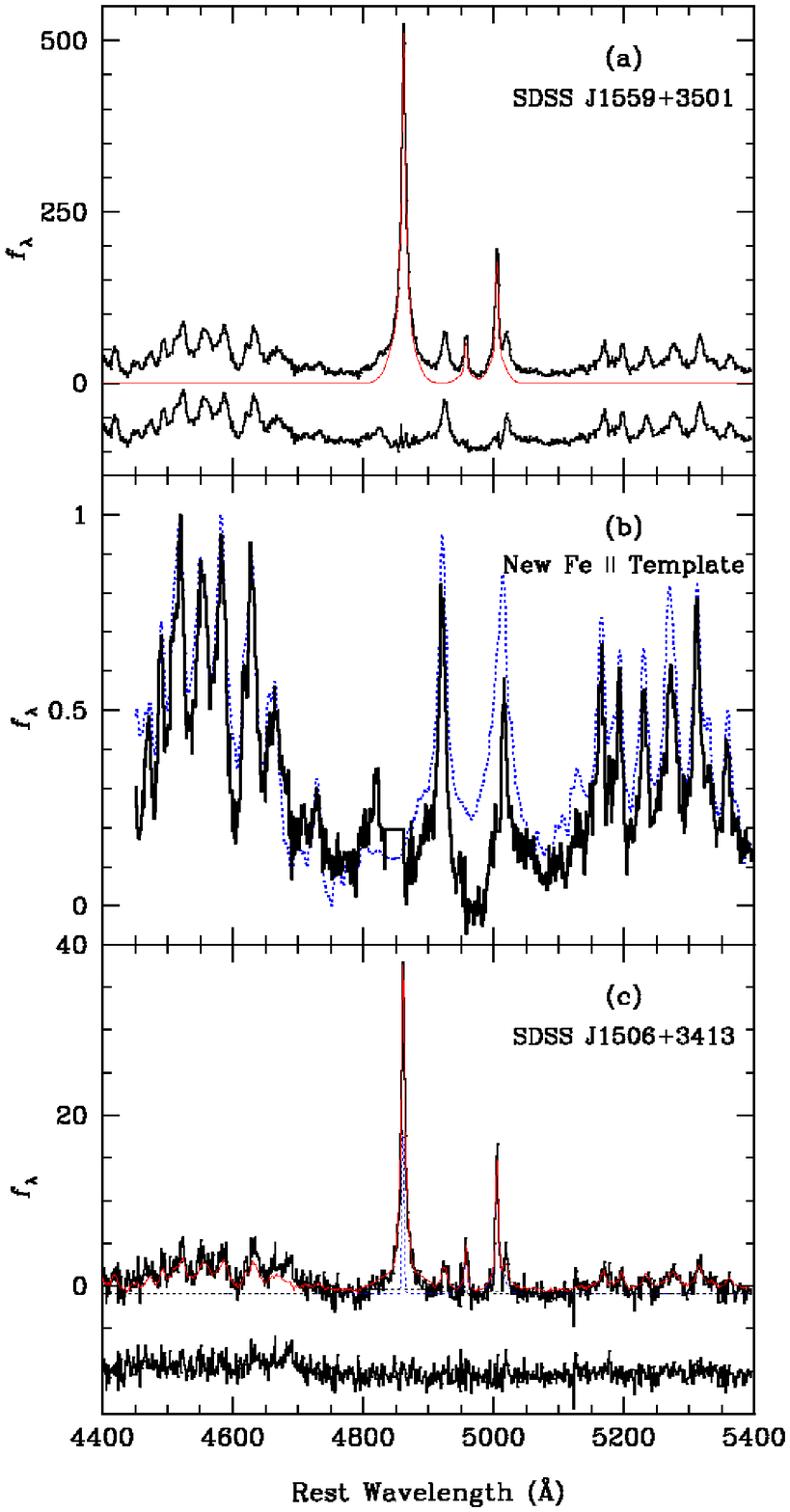,width=0.45\textwidth,keepaspectratio=true,angle=0}
\vskip -0mm 
\figcaption[]
{ ({\it a}) The \hbeta+[O {\tiny III}] fit to SDSS J155909.62$+$350147
used to create the Fe~{\tiny II} template shown in ({\it b}).  Shown
are the continuum-subtracted ({\it top}), best-fit ({\it red thin}) and
residual ({\it bottom}) spectra. Units are $10^{-17}$~\flamb.  ({\it
b}) A comparison of the standard Boroson \& Green 1992 1~Zw~{\tiny
I} Fe~{\tiny II} template ({\it blue dotted}) with that used in this work
({\it solid}).  Spectra have been normalized to unity.  The two
templates are very similar, although our template is clearly narrower,
and, as discussed in the text, the overall slope is somewhat
different.  ({\it c}) A sample fit to the \hbeta+[O {\tiny III}]
region.  Spectra in solid are the continuum-subtracted ({\it top}) and
residual ({\it bottom}) spectra. The total model is overplotted ({\it
red thin}), along with individual components ({\it blue dotted}).  
Units the same as ({\it a}) above. 
\label{fezw}}
\vskip 4mm
\noindent
we find in general very good agreement in
properties between the two samples.

\begin{figure*}
\vbox{ 
\vskip -0.1truein
\hskip 0.3in
\psfig{file=table_line_sh.epsi,width=0.85\textwidth,keepaspectratio=true,angle=0}
}
\end{figure*}

\subsection{Comparison with Dong et al. (2007a, b)}

Dong \etal\ (2007a) have presented a low-mass BH with a mass of $7
\times 10^4$~\msun.  This source was selected from the Fifth Data
Release of the SDSS, and so cannot be cross-checked with our sample.
However, the object has been serendipitously observed with \hst\ and
\xmm, as well as being detected with the R\"{o}entgen Satellite
(\rosat).  There is tentative evidence for variability in the X-ray
source ($L_{\rm X} =7 \times 10^{40}$~\lum) and the host galaxy is a disk
galaxy with $M_R = -17.8$ mag and a nuclear bar.  The same group has
identified a sample of 245 low-mass BHs from DR4, with a mass range of $5
\times 10^4 - 10^6$~\msun, and Eddington ratios ranging from 0.02 to
8, comparable to our sample (Dong \etal\ 2007b).  Samples selected
from two independent methods will provide important tests on the
limitations of our method.  Note, however, that the published masses 
are based on the radius-luminosity relation slope of Kaspi \etal\ (2005), 
and would increase by $\sim 0.3$ dex according to the formalism adopted in 
this paper.

\section{Sample Properties}

The final sample has a median redshift of 0.086 (0.099 if the \cs\
sample is excluded).  The original sample has a slightly lower median
$z = 0.08$ due in part to subsequent deep observations in the Southern
Strip (targeting quasars to a limiting magnitude of $i = 19.9$ mag
rather than $i=19.1$ mag).  In Figure 1 we show the distribution of
``virial'' BH mass and Eddington ratio for the sample, where
$L_{\mathrm{Edd}} \equiv 1.26 \times 10^{38}$~(\mbh/\msun) \lum.
Assuming that $L_{\rm{bol}} = 9.8$ \lf\ (McLure \& Dunlop 2004), in
terms of \lha\ our bolometric correction is $L_{\rm{bol}} = 2.34
\times 10^{44} (L_{\rm H\alpha} / 10^{42})^{0.86}$ \lum\ (Greene \& Ho
2005b).  The median BH mass is $\langle$\mbh$\rangle = 1.3 \times
10^6$~\msun.  With a median $\langle$\lledd$\rangle = 0.4$, we are
clearly dominated by sources radiating at substantial fractions of
their Eddington limits (Fig. 1).  This is not surprising; as discussed
by Greene \& Ho (2007b), we are only sensitive to the most luminous
BHs in this mass regime.

\subsection{Optical Spectral Properties}

The statistical power of our new enlarged sample provides new
constraints on the ensemble physical properties of both the radiating
BHs and their host galaxies.  We begin by placing the sources on the
two-dimensional diagnostic diagrams that in combination discriminate
between a stellar ionizing spectrum or one considerably harder
(e.g.,~Baldwin \etal\ 1981; Veilleux \& Osterbrock 1987).  These line
ratios are typically used to divide emission-line galaxies into
\ion{H}{2} galaxies and narrow-line AGNs (e.g.,~Ho \etal\ 1997a;
Kauffmann \etal\ 2003; Hao \etal\ 2005; Table 2).  In contrast, 
broad-line--selected objects are unbiased with regard to position on the
diagnostic diagrams, while at the same time we know that some fraction
of the ionizing continuum is contributed by an AGN.

We focus first on the \nii~$\lambda 6583$/\halpha\ diagnostic diagram
(Fig. 3{\it a}).  We do not see many objects in the lower-right
quadrant.  Objects there are low-ionization nuclear emission-line
region (LINER; Heckman 1980) sources and typically are highly
sub-Eddington AGNs (Ho 2004).  Compared to classical Seyfert galaxies,
which occupy the upper-right region of the diagnostic diagram, we see
that our sources span a broader range in both \nii/\halpha\ and
\oiii/\hbeta.  The decrease in \nii/\halpha\ is most easily explained
as a decrease in the gas-phase metallicity of these AGNs
(e.g.,~Kraemer \etal\ 1999; Groves \etal\ 2006). Since nitrogen is a
secondary element, \nii\ is particularly sensitive to changes in
metallicity for $> 0.1\, Z_{\sun}$.  In contrast, the \oiii/\hbeta\
ratio changes much more slowly, because as the metallicity decreases,
the temperature of the NLR increases correspondingly, which tends to
boost the \oiii\ strength (e.g.,~Groves \etal\ 2004a, 2006; see
metallicity tracks on Fig. 4).  Relatively low metallicities for this
sample are not unexpected, since local AGN hosts in general are known
to be massive, bulge-dominated galaxies (e.g.,~Ho \etal\ 1997b;
Kauffmann \etal\ 2003), while these low-mass BHs tend to live in
low-mass and correspondingly low-metallicity systems (\S 3.4; Tremonti
\etal\ 2004).

More striking is the apparent corresponding decrease in \oiii/\hbeta,
which as we have seen, cannot be ascribed to a decrease in
metallicity.  In fact, \oiii/\hbeta\ appears to decrease in concert
with \nii/\halpha, \sii/\halpha, and \oi~$\lambda 6300$/\halpha\
(Fig. 3{\it b, c}).  At the same time, there is a clear trend of
increasing \oii~$\lambda 3727$/\oiii\ line ratios as the
\oiii/\hbeta\ ratio decreases (Fig. 4).  Changes in a variety of NLR
conditions, including metallicity, electron density ($n_e$),
ionization parameter ($U$), ionizing spectral shape, and reddening
might be invoked to explain the observed line ratios, while particular
properties of either the AGN (e.g.,~luminosity or Eddington ratio) or
the host galaxy (e.g.,~star formation) may be responsible for the
changing conditions.  As we will argue, it is unlikely that a single
parameter is responsible for all of these trends.  Correlated errors
might also be expected to move line ratios in particular ways across
the diagnostic diagrams.  If, for instance, we are systematically
overestimating the strength of narrow \halpha\ in systems with weak
NLR emission, then we might expect correlated tracks as observed.
However, \oiii\ is never blended with broad \hbeta\ in these sources,
and thus \oiii\ errors are not correlated with the other lines, but
nevertheless the \oiii\ line strength spans a wide range relative to \hbeta.

\begin{figure*}
\vbox{ 
\vskip -0.1truein
\hskip 0.1in
\psfig{file=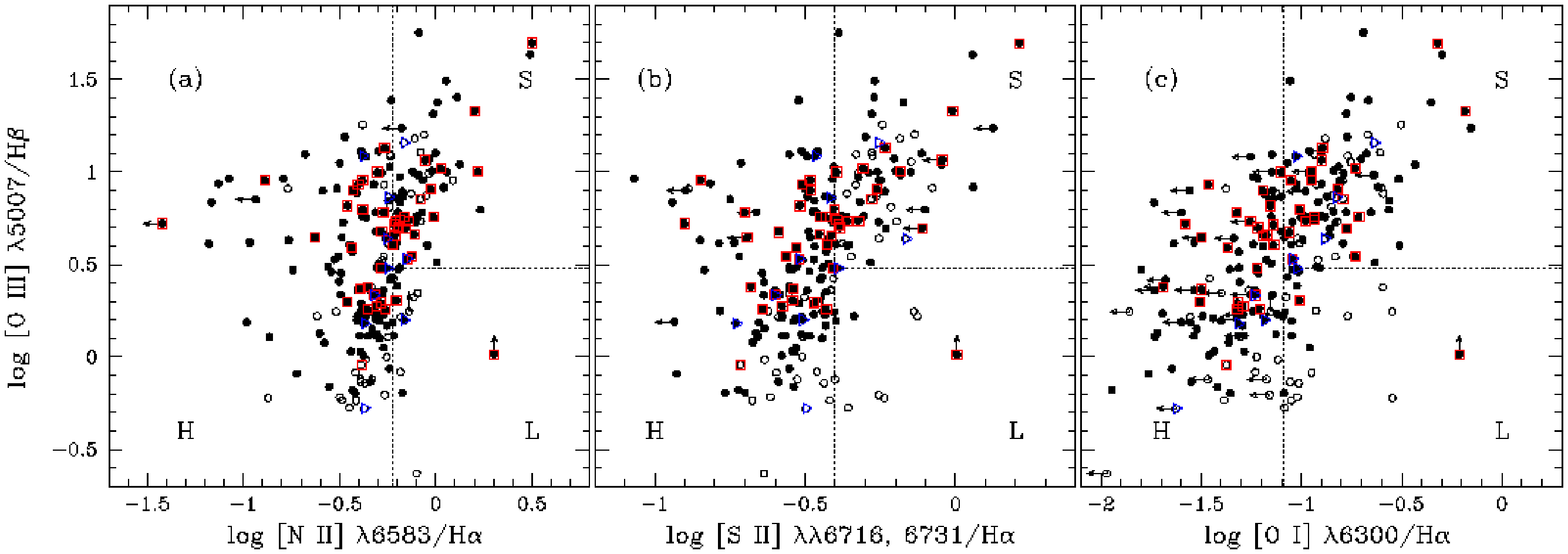,width=0.3\textwidth,keepaspectratio=true,angle=90}
}
\vskip -0mm
\figcaption[]{
Diagnostic diagrams plotting log [O~{\tiny III}] $\lambda$5007/\hbeta\
versus ({\it a}) log [N~{\tiny II}] $\lambda$6583/\halpha, ({\it b})
log [S~{\tiny II}] $\lambda\lambda$6716, 6731/\halpha, and ({\it c})
log [O~{\tiny I}] $\lambda$6300/\halpha.  The \cs\ sample is
shown in open symbols, and there are red boxes around \rosat\
detections and blue triangles around FIRST detections.  The line
ratios have not been corrected for reddening, but this should not
matter because of the small wavelength separation of the lines.  The
dotted lines mark the boundaries of the three main classes of
emission-line nuclei, according to the convention of Ho et
al. 1997a: ``H'' = H~{\tiny II} nuclei, ``S'' = Seyferts, and ``L''
= LINERs.
\label{diag}}
\end{figure*}

One might imagine that the intrinsic line ratios are all located in
the Seyfert locus, but that contamination from star formation spreads
out their positions in the diagnostic diagrams.  We perform a simple
thought experiment to demonstrate that there must be an intrinsic
spread in line ratios to reproduce the observed trends.  Since
star-forming galaxies follow a well-defined sequence in the diagnostic
diagram (bounded by the Kauffmann \etal\ 2003 relation shown Fig. 5),
we can easily investigate the type of bias that star formation would
contribute.  In Figure 5 we have chosen three fiducial AGN positions
in the diagnostic diagram.  Position A represents a typical
high-ionization system, while B has low \oiii/\hbeta\ and
\nii/\halpha\ and C represents a LINER.  To each AGN we add varying
amounts of star formation with line ratios drawn from the Kauffmann
line, and the \hbeta\ strength in star formation varying from half to
twice the AGN \hbeta\ luminosity.  The tracks (shown as dash-dot,
solid, and dashed lines for positions A, B, and C, respectively)
demonstrate that while star formation may increase the spread in
positions on the diagnostic diagram, it alone cannot move objects from
the high-ionization position A to the low-\oiii/\hbeta\ positions we
observe.  There must be a true spread in the intrinsic AGN line
ratios.  Furthermore, position C is disfavored, since only with a
large fraction of the light contributed by low-metallicity star
formation might we move the objects from position C toward position B.
If the objects in this sample follow the mass-metallicity relation of
Tremonti \etal\ (2004), with typical luminosities of $M_g \approx
-19.3$ (Table 3, \S 3.4), they would not, on average, have sub-solar
metallicity.

Given that the dispersion in line ratios must be intrinsic to the NLR,
we investigate various means to explain the observed line-ratio
distributions.  Ho \etal\ (1993) find that narrow-line AGNs display a
similar trend in the occupation of diagnostic diagrams, which they
ascribe to a trend of harder ionizing spectral shape at lower
luminosity.  Exactly this trend has been seen in the X-ray to optical
slopes of broad-line AGNs over a wide range in luminosity
(e.g.,~Strateva \etal\ 2005), including for a subset of the GH sample
(Greene \& Ho 2007a).  A correlation between low-luminosities and
harder spectral shape would lead to a correlation between AGN
luminosity and line ratios, but we see no trends between AGN
luminosity (as traced by \lha) and either \oiii/\hbeta\ or
\oi/\halpha.

We have already discussed the impact of varying metallicity above, and
since \nii/\halpha\ is most sensitive to its variation, metallicity is
unlikely to explain the correlated decrease in \sii/\halpha,
\oi/\halpha\ and \nii/\halpha.  Furthermore, metallicity variations
alone cannot explain the large observed range in \oii/\oiii\ ratios
(Fig. 4).  Changes in either ionization parameter or density might
explain the \oii/\oiii\ ratios; increases in either leads to a
relative increase in \oiii.  Indeed, electron densities inferred from
the ratio of the \sii\ lines show a significant correlation with
\oiii/\hbeta.  The Spearman rank correlation coefficient is $\rho =
0.296$, with a probability $P<10^{-4}$ that no correlation is present.
However, there is only a mildly significant trend between electron
density and \nii/\halpha\ ($\rho = 0.197,~P=0.003$) and no trend with
\sii/\halpha\ ($\rho = 0.124,~P=0.06$; as one might expect given their
lower critical densities).  On the other hand, changes in ionization
parameter move, for example,~\nii/\halpha\ vs. \oiii/\hbeta\ on tracks
perpendicular to those observed (e.g.,~ Groves \etal\ 2004a, b).  A
single parameter seems unable to reproduce all of the observed trends
simultaneously.  Presumably this is due, in part, to stratification in
the density and ionization parameter of the NLR with radius, which
leads to different typical conditions in the emission regions of
different lines (e.g.,~Filippenko \& Halpern 1984).

In summary, a single parameter cannot be invoked to explain the
observed positions in the diagnostic diagrams.  

\hskip -0.2in
\psfig{file=table_mags_sh.epsi,width=0.5\textwidth,keepaspectratio=true,angle=0}\vskip 4mm
\noindent
Although changing physical conditions (such as electron density)
within the NLR may be important, the objects lying close to the
star-forming locus may simply have an extremely weak NLR, such that
the majority of the narrow emission is coming from \ion{H}{2} regions.
A similar phenomenon, the ``vanishing'' NLR, has been seen in a
substantial fraction of higher redshift AGNs at high Eddington ratios
(e.g.,~Netzer \etal\ 2004, 2006).  If observed correlations between
NLR size and AGN luminosity (e.g.,~Bennert \etal\ 2002; Schmitt \etal\
2003) are extrapolated to the highest luminosities, then one would
predict unbound nebulosities; Netzer et al. (2004) suppose that the
NLRs in these systems grew so large that they were no longer bound to
their host galaxies.  The luminosities in our systems are
significantly lower, and the expected NLR sizes are $< 100$~pc if they
obey the NLR size-luminosity relation, but we cannot rule out that the
AGN actually expelled some large fraction of the surrounding ISM,
leading to overall weakness in the line intensities.

Finally, there is an apparent connection between \oiii/\hbeta\ and
Eddington ratio (e.g.,~Boroson 2002).  However, for these objects
there is no trend between \oiii/\hbeta(narrow) and \lledd\ ($\rho =
0.043$, $P = 0.5$) or \oiii\ equivalent width and \lledd\ ($\rho =
0.093$, $P = 0.2$).  These trends do not change when the {\it c}
sample is excluded.  On the other hand, although no physically
motivated explanation exists, low \oiii/\hbeta\ traditionally was a
defining characteristic of the class of objects known as narrow-line
Seyfert 1 galaxies (NLS1s; Osterbrock \& Pogge 1985), 

\psfig{file=table_xray_sh.epsi,width=0.4\textwidth,keepaspectratio=true,angle=0}
\vskip 4mm
\noindent
which are broad-line AGNs with relatively narrow broad-line widths.
We turn now to a general comparison of our objects with NLS1s.

\subsection{Comparison with NLS1s}

While NLS1s are an observationally defined class, a picture has
emerged, based on their strong \feii/\hbeta\ and weak \oiii/\hbeta\
line ratios, strong soft X-ray excess (e.g.,~Boller \etal\ 1996) and
X-ray variability (Leighly 1999), that NLS1s are in general low-mass
BHs (thus accounting for the relatively narrow broad lines) in a high
accretion state (e.g.,~Pounds \etal\ 1995).  Since the current sample is
selected based on BH mass (and indirectly \lledd), it represents a
very uniform, optically selected NLS1 sample.  In this section, we
highlight comparisons between the optical and broad spectral properties
of this and other NLS1 samples.

One of the supporting pieces of evidence that NLS1s are radiating
close to their Eddington limit comes from their observed high
\feii/\hbeta\ ratios (e.g.,~V{\'e}ron-Cetty \etal\ 2001; throughout
this section we refer to {\it total} [broad+narrow] \hbeta\
luminosity).  Using principle component analysis, Boroson \& Green
(1992) found that NLS1s lie at one extreme of their first component
(fondly known as Eigenvector 1), and later work has shown that
Eigenvector 1 properties, including high \feii/\hbeta\ and low
\oiii/\hbeta\ ratios, 

\hskip -0.2in
\psfig{file=o2o3.epsi,width=0.45\textwidth,keepaspectratio=true,angle=0}
\figcaption[]{
Tracks in [O {\tiny II}]/[O {\tiny III}] for different
metallicities, following Groves \etal\ 2006.  At a given
metallicity, the tracks span a range in ionization parameter \~{U}$_0
\equiv S_*/(c $\~{n}$_{\rm H})$, where $S_*$ is the number of ionizing
photons, and \~{n}$_{\rm H} = P_{\rm gas}/(k 10^4)$.  The model values are
log \~{U}$_0$=($-0.40,-1.03,-1.43,-1.88,-2.24,-2.57,-2.98$) and the tracks
are labeled at the low-\~{U}$_0$ end, as indicated with dashed lines.  It is
clear that metallicity alone cannot account for the spread in [O
{\tiny II}]/[O {\tiny III}] ratios we observe.  Open symbols indicate
the {\it c} sample.  
\label{o2o3}}
\vskip 5mm
\noindent
as well as soft X-ray excess and radio weakness,
tend to be correlated with Eddington ratio (e.g.,~Boroson 2002).  On
the other hand, GH found that their objects span a larger range in
both \feii\ and \oiii\ strengths relative to \hbeta\ than classic
NLS1s.  

The current sample has somewhat lower \feii/\hbeta\ strengths than GH
found, perhaps because there are objects with more significant galaxy
contamination in this sample.  Using the Kaplan-Meier product-limit
estimator, which accounts for upper limits (Feigelson \& Nelson 1985),
we find $\langle$\feii/\hbeta$\rangle = 0.57 \pm 0.03$ ($0.61 \pm
0.03$) with (without) the \cs\ sample, as compared to
$\langle$\feii/\hbeta$\rangle = 0.67 \pm 0.04$ for the 56 NLS1s
presented by V{\'e}ron-Cetty et al. (2001), and
$\langle$\feii/\hbeta$\rangle = 1.32 \pm 0.16$ for the original GH
sample.  The \oiii\ distributions are consistent with our previous
findings: $\langle$\oiii/\hbeta$\rangle = 2.6 \pm 0.3$ ($1.9 \pm
0.2$), while we found $\langle$\oiii/\hbeta$\rangle = 2.24 \pm 0.72$
for the original sample.  Although it has been customary to cite
\oiii/\hbeta(total), the ratio of \oiii/\hbeta(narrow) is more
straightforward to interpret, and we report it as well:
$\langle$\oiii/\hbeta(narrow)$\rangle =6.1 \pm 0.5$ ($6.5 \pm 0.6$).
In all other discussion, we report trends for \oiii/\hbeta(narrow).

More generally, trends in spectral properties with Eddington ratio may
be manifest in composite spectra.  We investigate four bins of
increasing Eddington ratio (similar to increasing bins in luminosity
over such a restricted range in BH mass; Fig. 6), and we have included
the {\it c} objects, which comprise the lowest \lledd\ bin.  To
compute the composite, each spectrum is normalized to the median value
in the spectral regions 4100-4800 \AA\ and 5100-5800 \AA, resampled
onto the same wavelength grid and then median combined.  Median
stacking preserves relative line ratios, but not the continuum shape
(e.g.,~Francis \etal\ 1991).  Although we do not explicitly weight the
spectra, they are each normalized to their respective continuum level,
which gives extra weight to faint sources.  Errors at each pixel are
calculated as the 68\% interquartile range of the included pixel

\hskip -0.1in
\psfig{file=starmix.epsi,width=0.45\textwidth,keepaspectratio=true,angle=0}
\vskip -0mm
\figcaption[]{
A diagnostic diagram highlighting how different effects move points in
the parameter space.  For pure AGNs with line ratios originating at
positions A, B, and C, we have mixed star-forming galaxies with line
ratios along the Kauffmann \etal\ 2003 line ({\it blue dotted}),
assuming that the \hbeta\ strength from photoionization by starlight
ranges from half (upper track; A in long dash-dot, B in solid, C in
long dash) to twice (lower track) that by the AGN.  The thin arrows
indicate the trajectory of increasing contributions from star
formation.  Also shown are photoionization models from Groves \etal\
2006.  Each track covers a single metallicity and the same range of
ionization parameters as Figure 4 above.  While mixing stellar and
AGN-excited line ratios may broaden the locus of points in the
diagnostic diagrams, it is clear that the intrinsic line ratios do not
all originate at position A.
\label{starmix}}
\vskip 5mm
\noindent
values, following, e.g.,~Fine \etal\ (2006).  Note that this procedure
is considerably simpler than that of, e.g.,~Francis \etal\ (1991),
since the redshifts measured by SDSS are highly reliable and the
spectral coverage and depth are so uniform for the majority of the
sample.  The clearest trend as we move to higher Eddington ratio is
the decreasing importance of the galaxy continuum.  At the same time,
the \feii\ features are much more apparent in the higher Eddington
ratio bins.  This appears to be in keeping with our expectations from
Eigenvector 1.  On the other hand, using the original, unbinned data,
we do not find significant evidence for a correlation between
\feii/\hbeta\ and \lledd\ ($\rho = -0.07$, $P = 0.3$).  There is a
decrease in the \oiii/\hbeta(total) ratio with Eddington ratio in the
composite spectra, which ranges from $3.6 \pm 1.6$ in the
lowest \lledd\ bin to $0.65 \pm 0.03$ (or $2.4 \pm 0.3$ when
considering only the narrow \hbeta).  Errors are dominated by the
variance in line strengths for individual objects comprising the
composite.

\begin{figure*}
\vbox{ 
\vskip -0.1truein
\hskip 0.in
\psfig{file=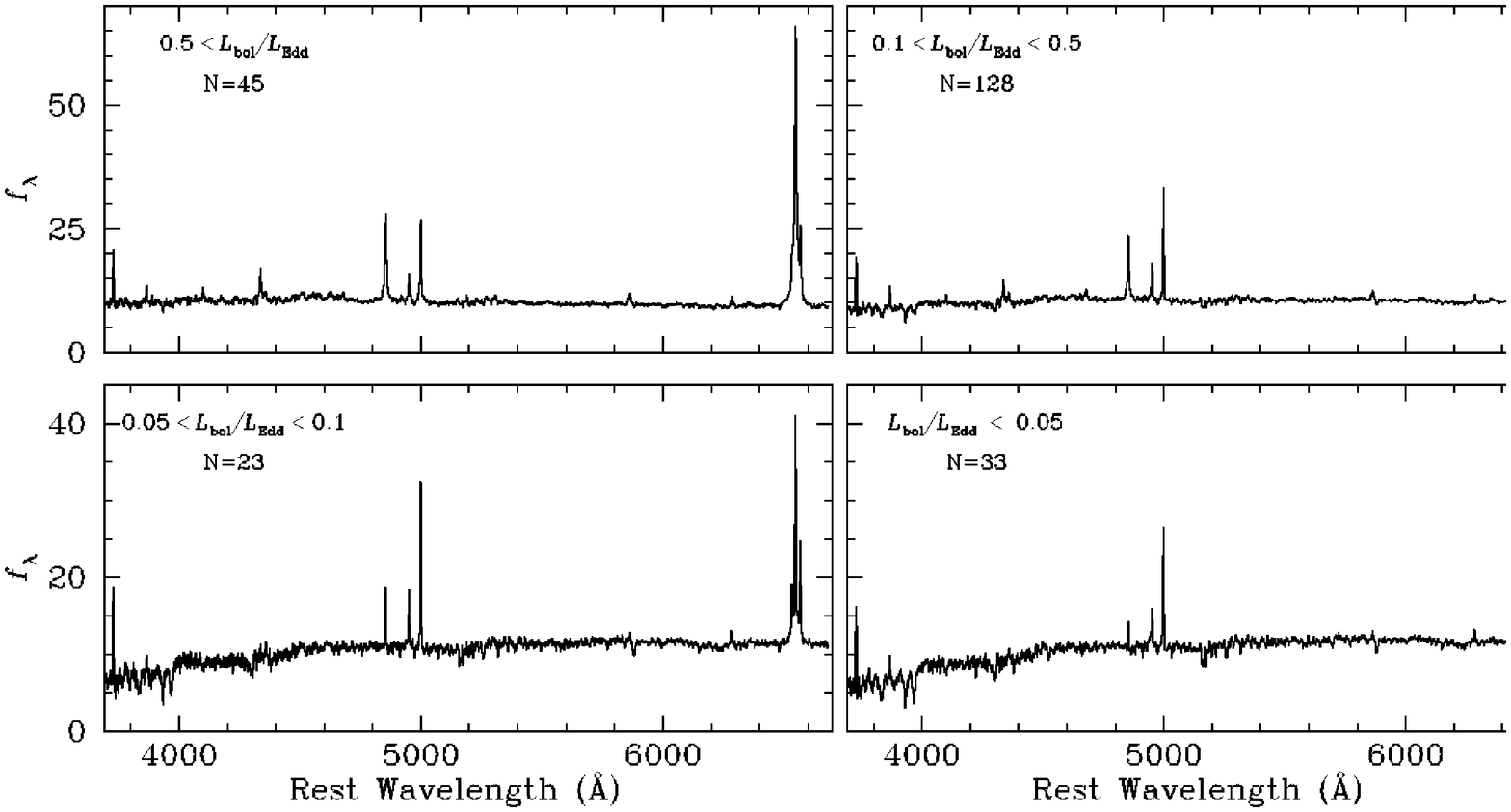,width=0.5\textwidth,keepaspectratio=true,angle=-90}
}
\vskip -0mm
\figcaption[]{
Composite spectra in bins of increasing Eddington ratio, as shown.
Each spectrum is the median combination of all spectra in the
indicated \lledd\ bin, with the total number of galaxies indicated.
Each spectrum is normalized to emission-line--free bands prior to
combination (see text for details).  Note the increasing strength of
the Fe {\tiny II} features and decreasing galaxy continuum strength
with increasing Eddington ratio.
\label{masshist}}
\end{figure*}

\subsection{Radio and X-ray Spectral Properties}

There is additional information about the accretion properties of the
sample in their broader spectral energy distributions.  NLS1s, in
general, are known for their strong soft X-ray excess, but Greene \&
Ho (2007a; see also Williams \etal\ 2004) have found that objects
selected purely on the basis of mass are not as extreme in this regard
as classic NLS1s.  In this case, using 5~ks observations of the 10
nearest GH objects, they found that the objects had very typical
0.5--2 keV power-law slopes of $\Gamma_s \approx 1.8$ [where $N(E)
\propto E^{-\Gamma_s}$].  Within the current sample, 55 are detected
by the \rosat\ All-Sky Survey (Voges \etal\ 1999), of which only two
are in the \cs\ sample, with fluxes ranging from $5.9 \times 10^{-14}$
to $3.4 \times 10^{-12}$ \flux.  Fluxes are derived from the \rosat\
count rates using {\tt WebPimms} (Mukai 1993) and assuming a slope of
$\Gamma_s = 2$.  In calculating uncertainties in the fluxes, 
$\Gamma_s$ is allowed to vary from 1 to 3, which is the
full range observed by Greene \& Ho (2007a).  The corresponding range
of 0.5--2 keV luminosities is $10^{40}$ to $7 \times 10^{43}$ \lum\
(Table 4).  Since inactive galaxies with optical luminosities similar
to those of the galaxies in our sample ($L_B \approx
10^9-10^{10}$~\lsun) are expected to have typical X-ray luminosities
of $\sim 10^{39}$~\lum\ (e.g.,~Fabbiano 1989), the AGN most likely
dominates the X-ray emission from the majority of these sources.

Customarily, the ratio of optical to X-ray luminosity is expressed in
terms of the slope of a supposed power-law between the optical and
X-ray bands, \aox$\equiv -0.3838\,$log($f_{\rm 2500 \AA}/f_{\rm 2
keV}$).  The flux density at 2500 \AA, $f_{\rm 2500 \AA}$, is
calculated using the measured \lha, the relation between \lha\ and
\lf\ from Greene \& Ho (2005b), and an assumed spectral slope of
$\beta = -1.56$ ($f_{\lambda} \propto \lambda^{-\beta}$; Vanden Berk
\etal\ 2001).  For our sample, the values lie in the range
$-1.5<$\aox$<-0.69$, with a median of $\langle$\aox$\rangle = -1.04$.
We would expect an \aox$\approx -1.2$ corresponding to the median
$l_{\rm 2500 \AA} = 6 \times 10^{27}$~ergs~s$^{-1}$~Hz$^{-1}$ (as was
seen for the GH objects observed with \chandra; Greene \& Ho 2007a).
While the observed median is slightly higher than one would expect
(e.g.,~Strateva \etal\ 2005; Steffen \etal\ 2006), it is within the
observed scatter, and given that these are the \rosat-detected members
of our sample, it is not surprising to find that they are X-ray
bright.  We see no obvious trends between \aox\ and \lledd\ or
\feii/\hbeta, but we caution that the dynamic range of this sample is
very limited, and that we do not have meaningful upper limits for the
X-ray--undetected sources.

Radio emission is a complementary probe of the accretion process, if
difficult to interpret cleanly.  In general, NLS1s are radio-quiet
(see review in Greene \etal\ 2006), where radio-quiet typically means
\rkel\ $< 10$ (\rkel\ $\equiv f_{\rm 6 cm}/f_{\rm 4400 \AA}$).  Weak
radio emission is part of Eigenvector 1 (e.g.,~Boroson \& Green 1992),
and mirrors the general tendency for BHs of all masses to be
radio-quiet when in high accretion states (e.g.,~McClintock \&
Remillard 2006).  Also, Best \etal\ (2005) find a strong preference
for radio sources to live in massive galaxies.  A low incidence of
radio activity in this sample is thus unsurprising. Eleven of the
objects are detected by the VLA\footnote{The VLA is operated by the
National Radio Astronomy Observatory, which is a facility of the
National Science Foundation, operated under cooperative agreement by
Associated Universities, Inc} FIRST survey at 20~cm although all but
18 objects are within the FIRST footprint (Becker \etal\ 1995.  The
radio sources at 20~cm are unresolved in all cases, with radio powers
that range from $2 \times 10^{21}$ to $9 \times 10^{22}$ \whz\ (Table
5).  Inactive star-forming galaxies can easily produce this level of
radio emission (e.g.,~Condon 1989), which need not be AGN-dominated.
The exception is GH10, which we have imaged with the VLA in
A-configuration at 6~cm (Greene et al. 2006).  We find that the source
remains completely unresolved, suggesting that the AGN provides the
dominant source of emission from this galaxy.

Of the objects with radio emission, four are in the \cs\ sample and
five others are radio-quiet, which translates into a radio-loud
fraction of $1\%$ (accounting for the 18 objects in the sample that
are outside of the FIRST footprint), 
to be compared with the range of $0\%-6\%$ for NLS1s in the literature
(Greene \etal\ 2006).  In Figure 7 the detected sources are placed on
the relation between radio power and \oiii\ luminosity observed for
local AGNs.  They are roughly consistent with the relation measured
for Seyfert galaxies (Ho \& Peng 2001), but they are systematically
higher than the fiducial relation.  In addition, we have retrieved
FIRST cut-outs around each non-detection.  We use these images to
derive upper limits on the radio luminosity (Table 5), and we also
stack the images in two bins divided at an \oiii\ luminosity of
\loiii=$3 \times 10^{40}$~\lum.  Because FIRST is a uniform survey, no
weighting is applied, but in 16 cases 

\hskip +0.5in
\psfig{file=table_radioall_sh.epsi,width=0.3\textwidth,keepaspectratio=true,angle=0}
\vskip 4mm
\noindent
there is a $\sim 2~\sigma$ detection in the inner five square pixels
(where $\sigma$ is measured in a box off the center of the cut-out).
We exclude these from our stacked images, although the answer changes
negligibly when they are included.  We do detect radio emission in
each stacked image, shown as crossed squares in Figure 7, also
consistent with the general relation.

\subsection{Host Galaxies}

Naively, our expectation is that lower-mass BHs will be found, in general,
in low-mass galaxies.  However, the tight observed relations between
BHs and galaxies pertain specifically to galaxy {\it bulges}.  
We are now in a position to test whether BH-bulge relations break down in 
low-mass galaxies.  The spectra provide an
estimate for the AGN luminosity (we prefer to use the \halpha\
luminosity because of its robustness in the presence of significant
galaxy contamination; Greene \& Ho 2005b), which we convert to
magnitudes using SDSS filter functions and assuming a spectral shape
of $\beta = -1.56$ (Vanden Berk \etal\ 2001).  If we assume a flatter
slope of $\beta = -1$, as suggested by GH, the median
magnitude difference is only 0.06 mag.  In addition to the power law, we
also include the observed strong emission 
lines (\hbeta, \oiii, \nii, and \halpha) in the calculation of the AGN
luminosity (on average, line emission accounts for $\langle g \rangle
\approx 0.08$ mag).  Once we have removed our best estimate for the
AGN luminosity from the SDSS Petrosian magnitudes, we calculate
$K$-corrections using the routines described in Blanton \& Roweis
(2007).  The distribution of host galaxy luminosities thus derived has
a median host galaxy luminosity of $\langle M_g \rangle = -19.3$ mag
(with or without the \cs\ sample) and is shown in Figure 8.  With this
method of AGN removal, we tend to underestimate the AGN luminosity due
to fiber losses, and thereby 

\hskip -0.1in
\psfig{file=rado3.epsi,width=0.45\textwidth,keepaspectratio=true,angle=0}
\vskip -0mm
\figcaption[]{
Adopted from Greene \etal\ 2006, this figure shows the relation
between \loiii\ and radio power for a large sample of AGNs.  Objects
from this paper are shown as open stars, while stacked luminosities
for undetected sources are shown as crossed squares and upper limits
from the original GH sample are shown as open triangles.  For clarity,
our data are plotted alone in the inset.  The solid line represents
the fit from Ho \& Peng 2001 to radio-quiet Seyfert galaxies and
Palomar-Green (PG; Schmidt \& Green 1983) quasars: \psix\ = (0.46
$\pm$ 0.15)~$L_{\mathrm{[O {\tiny III}]}}$ + (2.68 $\pm$ 6.21).  The
Ho \& Peng sample is overplotted with filled (radio-loud sources) and
open (radio-quiet sources) symbols.  PG quasars are shown as squares,
and Seyferts are shown as circles.  NGC 4395, with 
\loiii$=5.8 \times 10^{37}$~\lum\ (adjusted to a distance
of 4.2 Mpc; Thim et al. 2004) is highlighted as a semifilled circle.
\label{rad}}
\vskip 5mm
\noindent
(conservatively) overestimate the galaxy luminosity.  As a lower bound
on the galaxy luminosity, we use the SDSS point-spread function (PSF;
unresolved) magnitude as an alternate estimate of the AGN luminosity,
and then apply a $K$-correction to the corrected colors in the same
manner.  This method, on average, overestimates the AGN luminosity,
and results in a median host galaxy luminosity of $\langle M_g \rangle
= -19.0$ mag (Fig. 8, dotted histogram).  The true host luminosities
lie somewhere between these two bounds.  For reference, Blanton \etal\
(2003) find a characteristic luminosity of $M^*_g = -20.1$ mag at
$z=0.1$ (for our assumed cosmology), and thus our targets are $\sim
0.8-1.1$ mag fainter than $M^*_g$.

Morphological information is difficult to extract from the SDSS images
alone, due to their limited depth and spatial resolution, but it is
clear that BHs selected to have low masses are found in low-mass
galaxies.  However, if we restrict our attention to the 139 sources
with $z < 0.1$, we find $\sim 80$ clear disk galaxies (see GH for 
example SDSS images for the original sample).  We can also
calculate galaxy colors: $\langle g-r \rangle = 0.7$ mag for the
spectroscopic estimate of the AGN luminosity (0.6 mag for the PSF
method).  According to Fukugita \etal\ (1995), the host galaxies have
the colors of typical Sab galaxies ($g-r = 0.66$ mag).  \hst\
images of the GH sample indicate that there are actually two different
populations of host galaxies.  One is composed of late-type spiral
galaxies, typically with strong bars.  The other---the
majority---group is comprised of compact spheroids, not unlike POX 52,
the prototypical dwarf spheroidal galaxy hosting an AGN, which has red
colors ($B-V=0.8$ or $g-r \approx 0.67$ mag).  We do not yet know if
the spheroidal galaxies in the larger sample are red in general.

\subsection{Stellar Populations and Ongoing Star Formation}

It is tempting to imagine that the constant mass ratio observed
between BHs and bulges results from concurrent (or at least
orchestrated) star formation and AGN activity.  Indeed, various
theories have postulated that AGN activity is responsible for
truncating star formation (e.g.,~Springel \etal\ 2005).  While
characterizing the star formation rates of broad-line AGNs is
notoriously challenging, evidence for ongoing star formation has been
seen in some cases (e.g.,~Canalizo \& Stockton 2001).  With
narrow-line objects the galaxy is easier to study directly, and
Kauffmann \etal\ (2003) find a strong tendency for the most active
objects in their sample of SDSS narrow-line AGNs to contain a
significant post-starburst population.  Ho (2005) use the \oii\
doublet, which is intrinsically weak in high-ionization Seyfert
galaxies, to find that star formation is suppressed in luminous AGNs,
even when molecular gas is present (see also Kim \etal\ 2006); Ho
speculates that this may be a concrete manifestation of AGN feedback.
Our AGNs, while of low-mass and correspondingly low-luminosity,
nevertheless have the highest growth rates of any local AGNs because
of their high Eddington ratios (Greene \& Ho 2007b).  Unfortunately,
we do not have a large amount of information about the stellar
populations, both because of the AGN dominance and the low typical
signal-to-noise ratio of the spectra.

From the host galaxy colors, it is clear that recent star formation
has occurred in some fraction of the galaxies. A more compelling
argument for intermediate-age stellar populations comes from the
strength of the higher-order Balmer lines apparent in the stacked
spectra (Fig. 6).  Balmer absorption is present in all Eddington ratio
bins, but we cannot, from these data alone, quantify whether there is
any correlation between AGN luminosity and strength of Balmer
absorption.  We are wary to measure stellar population indices for
these objects without a clean measurement of the AGN luminosity, but
the absorption-line spectra (particularly in panels {\it c} and {\it
d} of Fig. 6) clearly contain a contribution from A stars.

Each of these clues only places constraints on star formation in the
past $\sim 10^9$~yr.  As always, it is far more challenging to find
evidence for ongoing star formation, since the emission from the AGN
itself generates nebular emission.  The \hst\ morphologies do suggest
that star formation is ongoing at least in the spiral arms of the disk
galaxies in the GH sample, but a more quantitative estimate is
desirable.  This was the motivation of Ho (2005) to use the
\oii/\oiii\ ratio as an indicator of concurrent star formation.
Unfortunately, as we have seen above, we do not have a strong handle
on the physical properties in the NLR, and therefore cannot uniquely
interpret the observed spread in \oii/\oiii\ ratio.  
There is a mild correlation between the \oii/\oiii\ ratio
and the $g-r$ color ($\rho = -0.3$, $P < 10^{-4}$; including or excluding 
the {\it c} sample), which one might expect if the \oii\ arises from
current star formation.  On the other hand, if the NLR conditions
correlate with Hubble type (not an unreasonable possibility), and
hence indirectly with galaxy color, a similar correlation may result.
With these data alone, we cannot place strong constraints on the
ongoing star formation in these systems, nor are constraints on star
formation alone sufficient to implicate a causal connection with bulge
growth.  We would like to determine whether there is an excess of star
formation in the AGN hosts as compared to inactive dwarf galaxies.  A
combination of high spatial resolution optical imaging with \hst\ and
broader spectral modeling of the host colors should be a first step
toward this goal.

\hskip 0.3in
\psfig{file=gal.epsi,width=0.35\textwidth,keepaspectratio=true,angle=0}
\vskip -0mm
\figcaption[]{
Distributions of $g$-band absolute magnitudes.  The magnitudes pertain
to the AGN alone ({\it top}), the host galaxy alone ({\it middle}),
and the entire system ({\it bottom}).  Filled histograms represent
objects below the detection threshold defined in Greene \& Ho (2007b;
the {\it c} sample).  The AGN contribution was estimated from \lha,
using the \lha--\lf\ relation of Greene \& Ho 2005b, and assuming a
spectral shape of $\beta = -1.56$ ($f_{\lambda} \propto
\lambda^{-\beta}$; Vanden Berk \etal\ 2001).  Alternatively, the AGN
magnitude may be estimated from the SDSS PSF magnitude, which leads to
the distribution shown in dotted lines in the middle panel.
\label{gal}}
\vskip 5mm

\section{Summary and Future Work}

We have increased the sample of low-mass BHs in galactic nuclei by
an order of magnitude (174+55 additional candidates).  In
general the sample properties are completely consistent with that of
the original GH sample.  The objects are radiating at $\sim$ one-third
of their Eddington limit, and many of them have \rosat\ counterparts.
On the other hand, as expected, very few have radio detections in the 
FIRST survey.  

Much like our previous sample, the current objects are found in
sub-$L^*$ galaxies, suggesting that indeed low-mass BHs are found in
low-mass stellar systems.  The colors and composite spectral
properties indicate that there are intermediate-age stellar
populations, but we have no direct constraints on ongoing star
formation.  Although the SDSS imaging is neither deep enough nor of
high enough angular resolution to determine the host galaxy structure,
we do have relevant external data for the original GH sample.  The
bulge stellar velocity dispersions presented in Barth \etal\ (2005)
show that the objects are consistent with the low-mass extrapolation
of the \msigma\ relation.  There is a suggestion, from Greene \& Ho
(2006), that the slope of the AGN \msigma\ relation flattens somewhat
at these low masses.

While the systems (at least to first order) obey the \msigma\
relation, they do not necessarily contain classical bulges.  NGC 4395
is an Sdm spiral galaxy (i.e.,~no bulge), while POX 52 is probably a
dwarf spheroidal.  As a result, the low-mass systems do not appear to
obey the low-mass extrapolation of the \mbh\--$L_{\rm bulge}$ relation
(e.g.,~Marconi \& Hunt 2003).  The total galaxy luminosities are
certainly larger than one would expect, based on their BH masses, and
this appears to be true even in the absence of disk components (Barth
\etal\ 2004; J.~E.~Greene et al. in preparation).  Therefore even near
their Eddington limits these BHs are not capable of providing as much
energy per unit host galaxy mass as more massive systems.  The
mystery, then, is why they appear to obey the same \msigma\ relation.

Along these lines, we should note that the revised radius-luminosity
relation slope (Bentz \etal\ 2006) substantially increases the virial
masses at low BH mass, which in turn increases the apparent flattening
in the \msigma\ relation slope at low mass previously reported by
Greene \& Ho (2006).  More insight should be gained into the reality
of this slope change once we have measured \sigmastar\ for the present
sample.  Ultimately, we wish to elucidate the degree to which the
slope and (crucially) intrinsic scatter of the \msigma\ relation has
any mass dependence (e.g.,~Robertson \etal\ 2006; Greene \& Ho 2006).

Indeed, there are a number of interesting applications to pursue with
the improved statistics afforded by this new, larger sample.  We plan
to examine the environments and clustering properties of these
objects, and see how they compare to inactive galaxies of similar mass
and color.  At the same time, we would like to constrain the stellar
populations of the host galaxies in a more quantitative fashion.  For
the GH sample, at least, this will be enabled by a combination of
\hst\ imaging and \spitzer\ IRS spectra.  Most immediately, we will
use the \hst\ imaging, in combination with stellar velocity dispersion
measurements, to place the GH sample on the fundamental plane.  In
combination, we will use these data to examine differences in the
evolution of low-mass systems as compared to supermassive BHs in
massive bulges.

\acknowledgements
We acknowledge the useful suggestions of the anonymous referee that
improved this work significantly.  We thank Michael Blanton for making
his $K$-correction code publicly available, and Brent Groves for kindly
providing his metallicity models prior to publication.  We thank the
entire SDSS team for providing the fantastic data products that made
this work possible.  Support for J.~E.~G. was provided by NASA through
Hubble Fellowship grant HF-01196 awarded by the Space Telescope
Science Institute, which is operated by the Association of
Universities for Research in Astronomy, Inc., for NASA, under contract
NAS 5-26555.  L.~C.~H. acknowledges support by the Carnegie
Institution of Washington and by NASA grant SAO 06700600.  Funding for
the SDSS has been provided by the Alfred P. Sloan Foundation, the
Participating Institutions, the National Science Foundation, the
U.S. Department of Energy, the National Aeronautics and Space
Administration, the Japanese Monbukagakusho, the Max Planck Society,
and the Higher Education Funding Council for England. The SDSS web
site is {\tt http://www.sdss.org/}.  This research has made use of
data obtained from the High Energy Astrophysics Science Archive
Research Center (HEASARC), provided by NASA's Goddard Space Flight
Center.

\appendix

\section{Virial Black Hole Mass Measurements}

The virial technique is quite indirect,
and its calibration is a matter of ongoing research.  Therefore, it is
worth describing our methodology in some detail.  In particular, we
discuss the potential systematic errors incurred by uncertainties in
the slope of the radius-luminosity relation on the one hand and the
derivation of robust gas velocity dispersions on the other.

At the present time, the radius-luminosity relation is calibrated
using the $\sim 30$ objects with BLR radii determined from
reverberation mapping.  Bentz \etal\ have used \hst\ to determine the
true AGN luminosity, as free as possible of host-galaxy contamination.
Their quoted radius-luminosity relation slope is significantly
shallower than that of Kaspi \etal\ (2005).  It is worth keeping in
mind the important caveat that the luminosities of the objects in that
study barely overlap with those in this work, and most are
significantly higher than those considered here.  While programs to
measure BLR radii for lower-luminosity sources are underway, at the
moment we have no choice but to adopt the results for the more
luminous sources.

As described in detail in Greene \& Ho (2005b), the line luminosity
provides a more robust measure of the AGN luminosity in the presence
of host galaxy contamination (or contamination from non-thermal jet
emission).  Therefore we have used the \lf-\lha\ relation from
Greene \& Ho (2005b) in combination with the revised radius-luminosity
relation from Bentz \etal\ (and uncertainties from M.~Bentz, private 
communication) to derive a final BH mass estimator:

\begin{equation}
\mbh = (3.0^{+0.6}_{-0.5}) \times 10^6 
\left(\frac{\lha}{10^{42}~{\rm erg~s^{-1}}} \right)^{0.45 \pm 0.03}\left(
\frac{\fwha}{10^{3}~\mathrm{\kms}} \right)^{2.06 \pm 0.06} \msun,
\end{equation}

\noindent
where we assume a spherical BLR ($f = 0.75$).  Here we are using the
FWHM of \halpha\ as a measure of the gas velocity dispersion.  In
principle the width of the broad emission lines represents the gas
kinematics, but in practice it is unclear how to derive the most
robust velocity dispersion measure.  It has long been clear that the
broad line shape depends systematically on other AGN properties
including line width (or BH mass) and luminosity (e.g.,~Sulentic
\etal\ 2000); in general, narrower lines have more Lorentzian shapes
(e.g.,~V{\'e}ron-Cetty \etal\ 2001).  If the geometry and kinematics
of the BLR change with Eddington ratio (as suggested by these
observations) then the assumption of a single geometrical factor $f$
breaks down.  In other words, the measured virial mass for a given BH
may vary as changes in the Eddington ratio lead to changes in the BLR
kinematics and thereby $f$.  An extreme example of this effect is 
reported for AGNs with double-peaked Balmer lines (Bian \etal\ 2007).

Without independent information it is impossible to determine $f$ as a
function of \lledd.  Luckily, the \msigma\ relation provides an
independent handle on the BH mass.  Especially for the relatively
low-luminosity AGNs considered at low redshift, it is safe to assume
that the systems intrinsically obey the \msigma\ relation, and adopt
the \msigma-derived mass as the true BH mass.  Both Onken \etal\
(2004) and Greene \& Ho (2006) find evidence for an average 
offset\footnote{Note that the Greene \& Ho measurement, while
including many more AGNs, is still dominated by the
reverberation-mapped sources, which have considerably smaller
systematic errors in their virial BH mass estimates.}  of $-0.26$ dex
from masses derived with $f=0.75$.  However, as argued
above, we expect that $f$ is not a constant, but rather depends on
\lledd.  While Greene \& Ho (2006) looked for evidence of a trend
between $f$ and \lledd, their dynamic range in both BH mass and
luminosity was too limited to draw robust conclusions.

Collin \etal\ (2006) perform an exhaustive study of the relation
between line shape and virial mass.  They use the Onken \etal\ sample
of 14 AGNs with both reverberation-mapped masses and \sigmastar\
measurements to investigate the merits of different measures of the
line width.  Specifically, they compare virial BH masses estimated
from the FWHM and the luminosity-weighted second moment of the
velocity profile ($\sigma$) with the \sigmastar-derived BH mass.  As
described above, they find that narrower lines are peaky (small values
of FWHM/$\sigma$) while broader lines are boxy (large values of
FWHM/$\sigma$).  As a result, the inferred $f$ may depend not only on
whether FWHM or $\sigma$ is used to measure line width, but also on
the line shape.  Using the comparison between virial mass and
\sigmastar, they derive a different value of $f$ for each subset of
objects (9 peaky and 5 boxy) using both FWHM and $\sigma$.  They claim
that while $f$ is constant across line samples when $\sigma$ is used
as the velocity dispersion indicator, the FWHM provides a biased
measure of the virial mass because $f$ is smaller for the boxy subset.
However, investigation of their table reveals that the values of $f$
derived from FWHM measurements are also consistent (within the quoted
errors) with a constant value.  Although the result may very well be
real, it is certainly not statistically significant at this time.

Moreover, we question the premise of Collin \etal\ that the best
measure of line width is the one that minimizes changes in $f$.
Rather, since the BLR geometry really does change with \lledd, one
should calibrate different $f$ values for different average line
shapes, when sufficient statistics exist to do so in a robust fashion.
Collin \etal\ present an additional piece of information that we
believe favors use of the FWHM over $\sigma$.  Because they are using
reverberation-mapped AGNs, many epochs of observations are available.
One may derive a so-called rms spectrum that isolates the variable
part of the line.  As argued in, e.g., Peterson \etal\ (2004), it is
preferable to measure the width of the variable component of the line,
which hopefully reflects the kinematics of gas at the measured radii.
As it turns out, the FWHM measured from the mean and rms spectra are
more self-consistent than the respective $\sigma$ measurements.  There
appears to be a very broad, non-variable component of the line which
biases $\sigma$ more strongly than FWHM (perhaps an optically thin
component; e.g.,~Shields \etal\ 1995).  While the very broad component
cannot be removed in single-epoch observations, at least the FWHM is
less biased by it.  Finally, it is quite clear from Figure 1 of Collin
\etal\ that the scatter in $\sigma$ measurements is considerably
larger than in the FWHM.  The same result was reported by Greene \& Ho
(2006).

Based on all of these considerations, the following choices were made 
in calculating the BH masses presented herein:

\begin{itemize}

\item
We use the radius-luminosity relation reported by Bentz \etal\ (2006).
Note that this differs from Greene \& Ho (2007b).

\item
We use the FWHM to determine the gas velocity dispersion.

\item
Because we believe $f$ is not a constant, but rather depends on
\lledd, and in order to facilitate comparison with the literature, we
do not apply an $f$ correction but use the fiducial assumption of a
spherical BLR ($f = 0.75$; e.g.,~Netzer 1990).

\end{itemize} 

\noindent
Obviously, virial masses remain highly uncertain, and potentially
systematically biased.  In addition to the many uncertainties
highlighted above, it is not yet clear the degree to which the BLR is
a flattened structure, which means that inclination effects cannot be
properly accounted for, even in a statistical sense.  We are hopeful
that an increased sample of reverberation-mapped sources
(e.g.,~Kollatschny 2003), combined with an increase in the
number of measured \sigmastar\ measurements in AGNs, will decrease the 
outstanding systematic errors in these masses in the near future.


\newpage
\begin{thebibliography}{}

\bibitem[]{}Adelman-McCarthy, J.~K., et al. 2006, \apjs, 162, 38 

\bibitem[]{}Baldwin, J.~A., Phillips, M.~M., \& Terlevich, R. 1981,
\pasp, 93, 5 

\bibitem[]{}Barth, A.~J., Greene, J.~E., \& Ho, L.~C. 2005, \apj, 619, L151 
 
\bibitem[]{}Barth, A.~J., Ho, L.~C., Rutledge, R.~E., 
\& Sargent, W.~L.~W. 2004, \apj, 607, 90 

\bibitem[]{}Becker, R.~H., White, R.~L., \& Helfand, D.~J. 1995, \apj,
450, 559

\bibitem[]{}Bennert, N., Falcke, H., Schulz, H., Wilson, A.~S., 
\& Wills, B.~J. 2002, \apj, 574, L105 

\bibitem[]{}Bentz, M.~C., Peterson, B.~M., Pogge, R.~W., Vestergaard, M., 
\& Onken, C.~A. 2006, \apj, 644, 133

\bibitem[]{}Best, P.~N., Kauffmann, 
G., Heckman, T.~M., Brinchmann, J., Charlot, S., Ivezi{\'c}, {\v Z}., \& 
White, S.~D.~M. 2005, \mnras, 362, 25 

\bibitem[]{}Bian, W.-H., Chen, Y.-M., Gu, Q.-S., \& Wang, J.-M. 2007, 
\apj, accepted (arXiv:0706.2473)

\bibitem[]{}Blanton, M. R., et al. 2003, \apj, 592, 819

\bibitem[]{}Blanton, M.~R., \& Roweis, S. 2007, \aj, 133, 734 

\bibitem[]{}Boller, Th., Brandt, W.~N., \& Fink, H. 1996, \aap, 305, 53 

\bibitem[]{}Boroson, T.~A. 2002, \apj, 565, 78

\bibitem[]{}Boroson, T.~A., \& Green, R.~F. 1992, \apjs, 80, 109 

\bibitem[]{}Canalizo, G., \& Stockton, A. 2001, \apj, 555, 719 

\bibitem[]{}Collin, S., Kawaguchi, T., Peterson, B.~M., \& Vestergaard, M. 
2006, \aap, 456, 75

\bibitem[]{}Condon, J.~J. 1989, \apj, 338, 13 

\bibitem[]{}Dong, X., et al. 2007a, \apj, 657, 700 

\bibitem[]{}Dong, X., Wang, T., Yuan, W., Zhou, H., Shan, H., Wang, 
H., Lu, H., \& Zhang, K. 2007b, in The Central Engine of Active Galactic 
Nuclei, ed. L. C. Ho \& J.-M. Wang (San Francisco: ASP), in press

\bibitem[]{}Eisenhauer, F., et al. 2005, \apj, 628, 246 

\bibitem[]{}Fabbiano, G. 1989, \araa, 27, 87 

\bibitem[]{}Favata, M., Hughes, S.~A., \& Holz, D.~E. 2004, \apj, 607, L5 

\bibitem[]{}Feigelson, E.~D., \& Nelson, P.~I. 1985, \apj, 293, 192 

\bibitem[]{}Ferrarese, L., \& Merritt, D. 2000, \apj, 539, L9

\bibitem[]{}Ferrarese, L., Pogge, R.~W., Peterson, B.~M., 
Merritt, D., Wandel, A.,\& Joseph, C.~L.\ 2001, \apj, 555, L79 

\bibitem[]{}Filippenko, A.~V., \& Halpern, J.~P. 1984, \apj, 285, 458 

\bibitem[]{}Filippenko, A.~V., \& Ho, L.~C. 2003, \apj, 588, L13 

\bibitem[]{}Filippenko, A.~V., \& Sargent, W.~L.~W. 1989, \apj, 342, L11 

\bibitem[]{}Fine, S., et al. 2006, \mnras, 373, 613 

\bibitem[]{}Francis, P.~J., Hewett, P.~C., Foltz, C.~B., Chaffee,
F.~H., Weymann, R.~J., \& Morris, S.~L.\ 1991, \apj, 373, 465

\bibitem[]{}Fukugita, M., Shimasaku, K., \& Ichikawa, T. 1995, \pasp, 107, 945 

\bibitem[]{}Gaskell, C.~M., \& Ferland, G.~J. 1984, \pasp, 96, 393 

\bibitem[]{}Gebhardt, K., et~al. 2000a, \apjl, 539, L13

\bibitem[]{}------. 2000b, \apjl, 543, L5

\bibitem[]{}------. 2001, \aj, 122, 2469 

\bibitem[]{}Gebhardt, K., Rich, R.~M., \& Ho, L.~C. 2002, \apj, 578, L41 
 
\bibitem[]{}------. 2005, \apj, 634, 1093 

\bibitem[]{}Ghez, A.~M., Salim, S., 
Hornstein, S.~D., Tanner, A., Lu, J.~R., Morris, M., Becklin, E.~E., \& 
Duch{\^e}ne, G. 2005, \apj, 620, 744 

\bibitem[]{}Greene, J.~E., \& Ho, L.~C. 2004, \apj, 610, 722

\bibitem[]{}------. 2005a, \apj, 627, 721 

\bibitem[]{}------. 2005b, \apj, 630, 122 

\bibitem[]{}------. 2006, \apj, 641, L21

\bibitem[]{}------. 2007a, \apj, 656, 84

\bibitem[]{}------. 2007b, \apj, accepted (astroph/0705.0020)

\bibitem[]{}Greene, J.~E., Ho, L.~C., \& Ulvestad, J.~S. 2006, \apj, 636, 56 

\bibitem[]{}Groves, B.~A., Dopita, M.~A., \& Sutherland, R.~S. 
2004a, \apjs, 153, 9 

\bibitem[]{}------. 2004b, \apjs, 153, 75 

\bibitem[]{}Groves, B.~A., Heckman, 
T.~M., \& Kauffmann, G. 2006, \mnras, 371, 1559 

\bibitem[]{}Halpern, J.~P., \& Steiner, J.~E. 1983, \apjl, 269, L37 

\bibitem[]{}Hao, L., et al. 2005, \aj, 129, 1783 

\bibitem[]{}Heckman, T.~M. 1980, \aap, 87, 152

\bibitem[]{}Ho, L.~C. 2005, \apj, 629, 680 

\bibitem[]{}------. 2004, in Carnegie Observatories Astrophysics Series, 
Vol. 1: Coevolution of Black Holes and Galaxies, 
ed. L. C. Ho (Cambridge: Cambridge Univ. Press), 292

\bibitem[]{}Ho, L.~C., Filippenko, A.~V., \& Sargent, W.~L.~W. 1997a,
\apjs, 112, 315

\bibitem[]{}------. 1997b, \apj, 487, 568

\bibitem[]{}Ho, L.~C., Filippenko, 
A.~V., Sargent, W.~L.~W., \& Peng, C.~Y. 1997c, \apjs, 112, 391

\bibitem[]{}Ho, L.~C.~\& Peng, C.~Y. 2001, \apj, 555, 650 

\bibitem[]{}Ho, L.~C., Shields, J.~C., 
\& Filippenko, A.~V. 1993, \apj, 410, 567 

\bibitem[]{}Hughes, S.~A. 2002, \mnras, 331, 805 

\bibitem[]{}Kaspi, S., Maoz, D., Netzer, H., Peterson, B.~M., Vestergaard, M., 
\& Jannuzi, B.~T. 2005, \apj, 629, 61 

\bibitem[]{}Kauffmann, G., et~al. 2003, \mnras, 346, 1055 

\bibitem[]{}Kim, M., Ho, L.~C., \& Im, M. 2006, \apj, 642, 702 

\bibitem[]{}Kollatschny, W. 2003, \aap, 407, 461 

\bibitem[]{}Kraemer, S.~B., Ho, L.~C., Crenshaw, D.~M., Shields,
  J.~C., \& Filippenko, A.~V. 1999, \apj, 520, 564 

\bibitem[]{}Krolik, J.~H. 2001, \apj, 551, 72 

\bibitem[]{}Kunth, D., Sargent, W.~L.~W., \& Bothun, G.~D. 1987, \aj, 93, 29 

\bibitem[]{}Leighly, K.~M. 1999, \apjs, 125, 297 

\bibitem[]{}Marconi, A., \& Hunt, L.~K. 2003, \apj, 589, L21 

\bibitem[]{}McClintock, J.~E., \& Remillard, R.~A.\ 2006, 
in Compact stellar X-ray
sources. Edited by Walter Lewin \& Michiel van der Klis. Cambridge
Astrophysics Series, No. 39. Cambridge, UK: Cambridge University Press

\bibitem[]{}McLure, R.~J., \& Dunlop, J.~S. 2004, \mnras, 352, 1390 

\bibitem[]{}Merritt, D., Milosavljevi{\'c}, M., Favata, M., Hughes, S.~A., 
\& Holz, D.~E. 2004, \apj, 607, L9 

\bibitem[]{}Meylan, G., Sarajedini, A., Jablonka, P., Djorgovski, S.~G., 
Bridges, T., \& Rich, R.~M. 2001, \aj, 122, 830 

\bibitem[]{}Mukai, K. 1993, Legacy, vol.~3, 3, 21 

\bibitem[]{}Nelson, C.~H., Green, R.~F., Bower, G., Gebhardt, K.,
\& Weistrop, D. 2004, \apj, 615, 652

\bibitem[]{}Netzer, H. 1990, in Active Galactic Nuclei, 
ed. R.~D.~Blandford, H.~Netzer, L.~Woltjer, T.~Courvoisier, \& M.~Mayor, 
(Berlin: Springer), 57 

\bibitem[]{}Netzer, H., Mainieri, 
V., Rosati, P., \& Trakhtenbrot, B. 2006, \aap, 453, 525 

\bibitem[]{}Netzer, H., Shemmer, O., 
Maiolino, R., Oliva, E., Croom, S., Corbett, E., \& di Fabrizio, L. 2004, 
\apj, 614, 558

\bibitem[]{}Onken, C.~A., Ferrarese, L., Merritt, D., Peterson, B.~M.,
Pogge, R.~W., Vestergaard, M., \& Wandel, A. 2004, \apj, 615, 645 

\bibitem[]{}Osterbrock, D.~E., \& Pogge, R.~W. 1985, \apj, 297, 166 

\bibitem[]{}Peterson, B.~M., et al. 2005, \apj, 632, 799 

\bibitem[]{}------. 2004, \apj, 613, 682

\bibitem[]{}Pooley, D., \& Rappaport, S. 2006, \apj, 644, L45 

\bibitem[]{}Pounds, K.~A., Done, C., \& Osborne, J.~P. 1995, \mnras,
277, L5

\bibitem[]{}Robertson, B., Hernquist, L., Cox, T.~J., Di Matteo, T., Hopkins,
P.~F., Martini, P., \& Springel, V. 2006, \apj, 641, 90

\bibitem[]{}Schmidt, M., \& Green, R.~F. 1983, \apj, 269, 352 

\bibitem[]{}Schmitt, H.~R., Donley, 
J.~L., Antonucci, R.~R.~J., Hutchings, J.~B., Kinney, A.~L., \& Pringle, 
J.~E. 2003, \apj, 597, 768 

\bibitem[]{}Shields, J.~C., Ferland, G.~J., \& Peterson, B.~M. 1995, 
\apj, 441, 507 

\bibitem[]{}Spergel, D.~N., et al. 2003, \apjs, 148, 175 

\bibitem[]{}Springel, V., Di Matteo, T., \& Hernquist, L. 2005, \apj, 620, L79 

\bibitem[]{}Steffen, A.~T., Strateva, I., Brandt, W.~N., Alexander, D.~M., 
Koekemoer, A.~M., Lehmer, B.~D., Schneider, D.~P., \& Vignali, C. 
2006, \aj, 131, 2826 

\bibitem[]{}Strateva, I.~V., Brandt, W.~N., Schneider, D.~P., 
Vanden Berk, D.~G., \& Vignali, C. 2005, \aj, 130, 387 

\bibitem[]{}Sulentic, J.~W., Zwitter, T., Marziani, P., \&
Dultzin-Hacyan, D. 2000, \apjl, 536, L5 

\bibitem[]{}Thim, F., Hoessel, J. G., Saha, A., Claver, J., Dolphin, A., \& 
Tammann, G. A. 2004, \aj, 127, 2322

\bibitem[]{}Tremaine, S., et al. 2002, \apj, 574, 740

\bibitem[]{}Tremonti, C.~A., et al. 2004, \apj, 613, 898 

\bibitem[]{}Ulvestad, J.~S., Greene, J.~E., \& Ho, L.~C. 2007, \apj, 661, L151 

\bibitem[]{}Ulvestad, J.~S., \& Ho, L.~C. 2001, \apj, 558, 561 

\bibitem[]{}Valluri, M., Ferrarese, L., Merritt, D., \& Joseph, C.~L. 
2005, \apj, 628, 137 

\bibitem[]{}Vanden Berk, D.~E., et al. 2001, \aj, 122, 549 

\bibitem[]{}Veilleux, S., \& Osterbrock, D.~E. 1987, \apjs, 63, 295

\bibitem[]{}V{\'e}ron-Cetty, M.-P., V{\'e}ron, P., 
\& Gon{\c c}alves, A.~C. 2001, \aap, 372, 730 

\bibitem[]{}Voges, W., et al. 1999, \aap, 349, 389 

\bibitem[]{}Williams, R.~J., Mathur, S., \& Pogge, R.~W.\ 2004, \apj, 610, 737 

\bibitem[]{}York, D.~G., et~al., 2000, \aj, 120, 1579

\end{thebibliography}
\end{document}